\documentclass[aps,twocolumn,showpacs,amsmath]{revtex4}
\usepackage{graphicx}

\newcommand{\bra}[1]{\langle #1 | \,}
\newcommand{\ket}[1]{\, | #1 \rangle}

\newcommand{\de}{\delta}

\newcommand{\bk}{b^{\dagger}}

\newcommand{\hn}{\hat{n}}

\newcommand{\Uc}{\mathcal{U}}
\newcommand{\be}{\begin{equation}}
\newcommand{\ee}{\end{equation}}
\newcommand{\bea}{\begin{eqnarray}}
\newcommand{\eea}{\end{eqnarray}}
\newcommand{\besa}{\begin{subequations}\begin{eqnarray}}
\newcommand{\eesa}{\end{eqnarray}\end{subequations}}

\begin{document}

\title{Two-particle states in the Hubbard model}

\author{Manuel Valiente}
\author{David Petrosyan} 
\affiliation{Institute of Electronic Structure \& Laser, FORTH, 
71110 Heraklion, Crete, Greece}

\date{\today}

\begin{abstract}
We consider a pair of bosonic particles in a one-dimensional 
tight-binding periodic potential described by the Hubbard model 
with attractive or repulsive on-site interaction. 
We derive explicit analytic expressions for the two-particle 
states, which can be classified as 
(i) scattering states of asymptotically free particles, and 
(ii) interaction-bound dimer states. Our results provide 
a very transparent framework to understand the properties 
of interacting pairs of particles in a lattice. 
\end{abstract}


\maketitle

%Introduction
Many-body quantum systems in spatially-periodic potentials were 
studied since the early days of quantum theory \cite{SolStPh}.
Several simple but powerful models have been invented and successfully 
employed for the description of various condensed matter systems, 
including the Heisenberg spin and Hubbard models. Lately, 
the Hubbard model has been attracting renewed attention prompted 
by the remarkable experimental progress in cooling and trapping 
bosonic and fermionic atoms in optical lattices \cite{OptLatRev}.
The relevant parameters of this system can be controlled with  
high precision and can be tuned to implement the Hubbard model 
with unprecedented accuracy \cite{optlattMI}. 

Recently, Winkler {\it et al.} \cite{KWEtALPZ} have observed 
a remarkable Hubbard model phenomenon: A repulsive on-site 
interaction between a pair of bosonic atoms in an optical lattice 
can bind them together into an effective, dynamically stable ``dimer''
\cite{molmer,PSAF,MVDP}. Here we revisit the problem of two bosonic 
particles in a 1D (one-dimensional) periodic potential described by 
the Hubbard model. We consider both cases of attractive and repulsive
on-site interactions between the particles and derive simple, exact 
analytic solutions for the two-particle problem. One type of 
these solutions corresponds to the scattering states of 
asymptotically free particles, while the other describes the
interaction-bound dimer states with energies below or above 
the energy band of the scattering states \cite{molmer}. 

We note that the solution of the Schr\"odinger equation for the 
two-particle problem in a lattice can be derived using the Green 
function approach \cite{KWEtALPZ,molmer}, which although being 
more general, requires laborious calculations often accompanied
by approximations for obtaining explicit analytic results. 
In contrast, in what follows we give an extremely simple
and straightforward derivation of the energies and exact 
analytic wavefunctions for the two-particles states in a lattice.

%Bose-Hubbard Hamiltonian
The Hamiltonian of the problem is given by 
\be
H = -J \sum_{j} (\bk_j b_{j+1} + \bk_{j+1} b_j ) 
+ \frac{U}{2} \sum_{j} \hn_j (\hn_j-1), \label{BHHam}
\ee
where $\bk_{j}$ ($b_{j}$) is the bosonic creation (annihilation) operator 
and $\hn_j = \bk_{j} b_{j}$ the number operator for site $j$, 
$J (> 0)$ is the tunnel coupling between adjacent sites, and $U$ is 
the on-site interaction. A standard basis $\{\ket{n_j}\}$ for 
Hamiltonian~(\ref{BHHam}) is composed of the eigenstates 
$\ket{n_j} \equiv \frac{1}{\sqrt{n !}} (\bk_j)^{n} \ket{\mathrm{vac}}$
of operator $\hn_j$ whose eigenvalues $n = 0,1,2,\ldots$ denote the 
number of bosonic particles at site $j$, and 
$\ket{\mathrm{vac}} \equiv \ket{\{ 0_j \}}$ is the vacuum state.

%Single particle
For a single particle in the periodic potential, the last term on 
the right-hand side of Eq.~(\ref{BHHam}) does not play a role.
The Hubbard Hamiltonian then takes the form
\be
H^{(1)} = -J \sum_{j} (\ket{x_j}\bra{x_{j+1}} + \ket{x_{j+1}} \bra{x_j} ) 
, \label{HHamx}
\ee
where $\ket{x_j} = \ket{1_j}$ denotes the state with a single
particle at position $x_j \equiv d j$ of the $j$th lattice site,
with $d$ the lattice constant. Expanding the single-particle 
state vector as $\ket{\psi} = \sum_j \psi(x_j) \ket{x_j}$, the 
stationary Schr\"odinger equation $H^{(1)} \ket{\psi} = E^{(1)} \ket{\psi}$ 
reduces to the difference equation
\be
-J \big[ \psi(x_{j - 1} ) + \psi(x_{j + 1}) \big] = E^{(1)} \, \psi(x_j)  
\ee
for the wavefunction $\psi(x_j)$. Taking 
$\psi(x) = \psi_q(x) \equiv \exp(i q x)$ immediately yields 
the well-known result
\be
-2J \cos(q d) \, \psi_q(x_j) = E_q^{(1)} \, \psi_q(x_j) ,
\ee
which implies that the discrete plane waves $\psi_q(x_j) = \exp(i q d j)$
are the eigenfunctions of Hamiltonian (\ref{HHamx}), with the corresponding
eigenenergies $E_q^{(1)} = -2J \cos(q d)$, forming a Bloch band of width 
$4J$ \cite{SolStPh}. The effective mass $m^*$ of the particle with 
quasi-momentum $q$ close to 0 is then given by the usual expression
\be
m^* = \hbar^2 \left[\frac{\partial^2 E_q^{(1)}}{\partial q^2 } \right]^{-1}_{q=0} 
= \frac{\hbar^2}{2J d^2}. \label{effmass}
\ee

%Two particles
For two particles in a one-dimensional lattice, the Hubbard Hamiltonian 
can be recast in the explicit form
\bea
H^{(2)} &=& -J \sum_{j} (\ket{x_j}\bra{x_{j+1}} + \ket{x_{j+1}} \bra{x_j} ) 
\nonumber \\ & &
-J \sum_{j} (\ket{y_j}\bra{y_{j+1}} + \ket{y_{j+1}} \bra{y_j} )
\nonumber \\ & &
+ U \sum_{j} \ket{x_j, y_j} \bra{x_j, y_j} , \label{HHamxy}
\eea
while the two-particle state vector can be expanded in terms of 
the non-symmetrized basis $\{\ket{x_j , y_{j'}}\}$ as
\be
\ket{\Psi} = \sum_{j,j'} \Psi(x_j,y_{j'}) \ket{x_j , y_{j'}} , 
\ee
where $y_j \equiv d j$. Clearly, the standard (symmetrized) bosonic 
basis is related to the non-symmetrized basis via the transformation 
$\ket{2_j} = \ket{x_j, y_{j}}$ and $\ket{1_j,1_{j'}} = \frac{1}{\sqrt{2}}
(\ket{x_j, y_{j'}} + \ket{y_j, x_{j'}})$ ($j\neq j'$). The stationary 
Schr\"odinger equation $H^{(2)} \ket{\Psi} = E^{(2)} \ket{\Psi}$
is now equivalent to the recurrence relation
\bea
&& -J \big[\Psi(x_{j - 1},y_{j'}) + \Psi(x_{j + 1}, y_{j'}) 
\nonumber \\ && \qquad 
+ \Psi(x_{j},y_{j'-1}) +  \Psi(x_{j}, y_{j'+1}) \big] 
\nonumber \\ && \qquad \qquad
+ U \de_{jj'} \Psi(x_{j },y_{j'}) = E^{(2)} \, \Psi(x_{j },y_{j'}) .   
\label{tpwfxy}
\eea
Define the center of mass $R = \frac{1}{2}(x + y)$ and 
relative $r = x - y$ coordinates. In terms of the new variables, 
the two-particle wavefunction can be separated as 
\be
\Psi(x,y) = e^{i K R} \, \psi_K (r) ,
\ee
where the relative coordinate wavefunction $\psi_K (r)$ depends now
on the center-of-mass quasi-momentum $K \in [-\pi/d,\pi/d]$ as a 
continuous parameter. Equation (\ref{tpwfxy}) then yields
\bea
-J_K \big[ \psi_K (r_{i-1}) + \psi_K (r_{i+1}) \big] 
+ U \de_{r0} \psi_K (r_i) \qquad
&& \nonumber \\ 
= E_K^{(2)} \, \psi_K (r_i)  , & & \label{tpwfr}
\eea
with $ J_K \equiv 2 J \cos(Kd/2)$ and $r_i = d i$ ($i = j - j'$).

\begin{figure}[ht]
\includegraphics[width=0.48\textwidth]{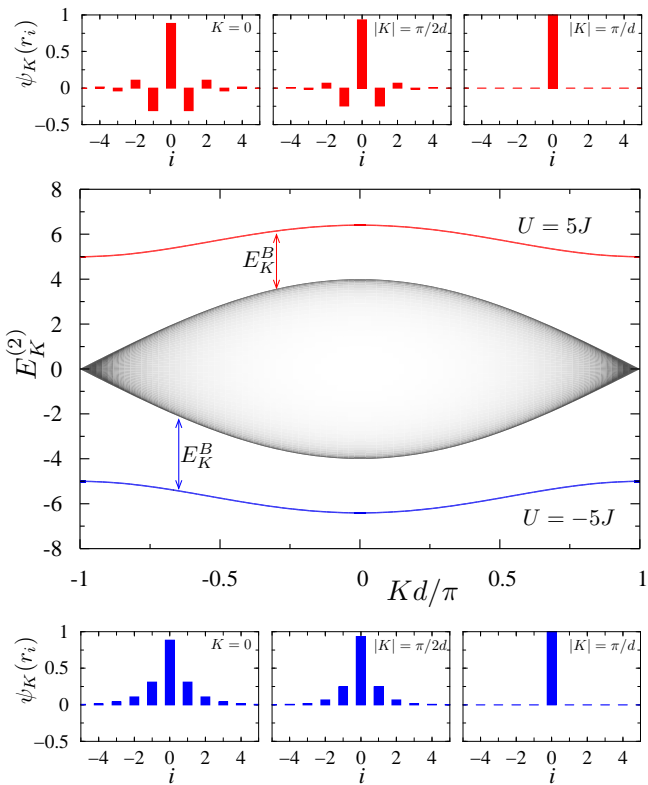}
\caption{Energies $E_{K}^{(2)}$ versus the center-of-mass momentum $K$
for a pair of bosons in a 1D lattice described by the Hubbard model.
The continuum spectrum corresponds to energies~(\ref{scatcont})
of the scattering states, with the shading proportional to the 
density of states (\ref{rhoEK}). The (blue) line below and
the (red) line above the scattering band are, respectively, 
the energies of the attractively-bound dimer with $U = - 5J$
and the repulsively-bound dimer with $U = 5J$. Their relative
coordinate wavefunctions (\ref{psiadim}) and (\ref{psirdim}) at 
$|K| = 0,\pi/2d,\pi/d$ are shown on the bottom and the top panels.}
\label{fig:drpwf}
\end{figure}

%Two particles: scattering solutions
(i) Consider first the scattering solutions for a pair of asymptotically 
free particles \cite{KWEtALPZ,molmer}. Since the action of the short-range
scattering potential $U \delta_{r0}$ amounts to a unitary phase-shift
(see below), the spectrum of such solutions is given by the sum 
of the spectra for two free particles $x$ and $y$ with momenta 
$q_x = K/2 + k$ and $q_y = K/2 - k$:
\bea
E_{K,k}^{(2)} = E_{q_x}^{(1)} + E_{q_y}^{(1)} &=& -4 J \cos(Kd/2) \cos(kd) 
\nonumber \\ 
&=& -2 J_K \cos(kd) . \label{scatcont}
\eea
Obviously, this result also follows directly from Eq.~(\ref{tpwfr}), 
with $U=0$, upon substitution of the plane waves 
$\psi_K (r) = \psi_{K,k}^0 (r) = \exp(\pm i k r)$ \cite{molmer}.
For a given value of the center-of-mass momentum $K$, and thereby $J_K$, 
the lowest  $E_{K,0}^{(2)} =  -2 J_K$ and highest $E_{K,\pi}^{(2)} =  2 J_K$
energy states of a pair of asymptotically free particles correspond, 
respectively, to the relative momenta $k \to 0$ and $k \to \pi/d$.  
The energy spectrum of Eq.~(\ref{scatcont}), and the corresponding
density of states 
\be
\rho(E,K) = \frac{L}{2 \pi} \frac{\partial k}{\partial E} 
= \frac{L}{2 \pi d} \frac{1}{\sqrt{[4 J \cos(Kd/2)]^2 - E^2}} , \label{rhoEK}
\ee
with $L$ a quantization length, are shown in Fig. \ref{fig:drpwf}.

We can now derive an explicit expression for the wavefunction of the 
scattering states. Substituting $E_{K,k}^{(2)}$ of (\ref{scatcont}) into 
Eq.~(\ref{tpwfr}), setting $\psi_{K,k} (0) = C$ and using the bosonic 
symmetry $\psi_{K,k} (r) = \psi_{K,k} (-r)$, after little algebra we obtain
\[
\psi_{K,k} (r_i) = C \cos(k r_i) + C \frac{U \csc (k d)}{2 J_K} 
\sin(k |r_i|).
\]
In terms of the plane waves undergoing the scattering phase shift 
$\de_{K,k}$, the wavefunction reads
\bea
\psi_{K,k} (r_i) &=& e^{i k |r_i|} e^{i \de_{K,k}} 
+ e^{-i k |r_i|} e^{- i \de_{K,k}} , \label{psiscat} \\ 
\tan(\de_{K,k}) &=& - \frac{U \csc (k d)}{2 J_K} . \nonumber
\eea
In the limit of $U \to 0$, this reduces to the trivial solution
$\psi_{K,k} (r_i)  = \cos(k r_i)$ for two non-interacting bosons,
while in the opposite limit of $U/J_K \to \pm \infty$, Eq. (\ref{psiscat})
yields the fermionized solution $\psi_{K,k} (r_i) = \sin (k |r_i|)$.
At the edges of the scattering band $k \to 0,\pi/d$, we can define 
a generalized 1D scattering length $a_K$ via \cite{molmer}
\be
a_K = - \lim_{k \to 0} \frac{\partial \de_{K,k}}{\partial k}
= - \frac{2 d J_K}{U} ,
\ee
which is thus positive for $U<0$ and negative for $U>0$, as can be 
intuitively understood on the basis of perturbation theory \cite{PSAF}. 
Thus, in the case of $U<0$, a pair of co-localised particles 
(forming a dimer discussed below) has an energy below that 
of a pair of unbound particles, causing therefore an effective 
repulsion between the unbound particles by pushing their energy upwards.
Conversely, the energy of the repulsively bound pair $U>0$ is 
above the energy of the unbound pair of particles, which
lowers their energy causing an effective attraction.

%Two particles: interaction-bound solutions
(ii) The on-site interaction $U \neq 0$ can bind the two bosonic particles 
together into a close dimer \cite{KWEtALPZ,molmer,PSAF} whose energy 
is above ($U > 0$) or below ($U < 0$) the continuum (\ref{scatcont})
of scattering states. We now derive simple analytic solutions for 
the bound dimer states. For $|K| = \pi/d$ ($J_K =0$), Eq.~(\ref{tpwfr}) 
immediately yields $E_{\pi/d}^{(2)} = U$ and
\be
\psi_{\pi/d} (0) = 1 , \qquad \psi_{\pi/d} (r_i \neq 0) = 0 .
\ee
For $K \in (-\pi/d,\pi/d)$,  using the exponential ansatz 
$\psi_K (r_i) = C \alpha_K ^{|i|}$, where $C$ is a normalization constant,
from Eq.~(\ref{tpwfr}) we have 
\besa
-2 J_K \alpha_K + U &=& E_K^{(2)} , \\ 
-J_K \frac{\alpha_K^{|i|-1} + \alpha_K^{|i|+1}}{\alpha_K^{|i|}}
&=& E_K^{(2)} ,
\eesa
where the unknowns are $E_K^{(2)}$ and $\alpha_K$. 
The solutions for $\alpha_K$ are given by
\[
\alpha_K = \Uc_K \pm \sqrt{\Uc_K^2 + 1}, 
\quad \Uc_K \equiv \frac{U}{2 J_K}.
\]
A normalizable bound state corresponds to $|\alpha_K| < 1$, which exists 
for any $U \neq 0$. Thus, for attractive interaction, $U < 0$, we have
\[
\alpha_K = \sqrt{\Uc_K^2 + 1} - |\Uc_K| > 0, 
\]
with which for the energy and normalized relative coordinate
wavefunction we obtain 
\besa
E_K^{(2)} &=& - \sqrt{U^2 + 4 J_K^2}, \label{Ekadim} \\
\psi_K (r_i) &=&  \frac{\sqrt{|\Uc_K|}}{\sqrt[4]{\Uc_K^2 + 1}} \, 
\Big( \sqrt{\Uc_K^2 + 1} - |\Uc_K| \Big)^{|i|}. \label{psiadim}
\eesa 
The lowest energy of an attractively-bound dimer is 
$E_0^{(2)} = - \sqrt{U^2 + 16 J^2}$ attained at $K=0$, 
while the highest energy is $E_{\pi/d}^{(2)} = -|U| = U$
at $K= \pm \pi/d$. We can define the binding energy for a dimer 
with momentum $K$ as the difference between the energy of the 
bound pair $E_K^{(2)}$ and the energy at the bottom of the scattering band 
for an unbound pair $E_{K,0}^{(2)} = -2 J_K$ 
(see Fig. \ref{fig:drpwf}, central panel),
\[
E_K^{B} \equiv E_K^{(2)} - E_{K,0}^{(2)} = 2 J_K - \sqrt{U^2 + 4 J_K^2} < 0.
\]

For repulsive interaction, $U > 0$, we have
\[
\alpha_K = \Uc_K - \sqrt{\Uc_K^2 + 1} < 0,
\]
and the dimer energy and wavefunction are given by
\besa
E_K^{(2)} &=& \sqrt{U^2 + 4 J_K^2}, \label{Ekrdim} \\
\psi_K (r_i) &=&  \frac{\sqrt{\Uc_K}}{\sqrt[4]{\Uc_K^2 + 1}} \, 
\Big(\Uc_K - \sqrt{\Uc_K^2 + 1} \Big)^{|i|}.  \label{psirdim}
\eesa
The lowest and highest energies of a repulsively-bound dimer are
given by $E_{\pi/d}^{(2)} = |U| = U$ and $E_0^{(2)} = \sqrt{U^2 + 16 J^2}$,
attained, respectively, at $K= \pm \pi/d$ and $K= 0$.
The dimer binding energy, counted from the top of the scattering band
$E_{K,\pi}^{(2)} = 2 J_K$ (see Fig. \ref{fig:drpwf}), is now positive, given by
\[
E_K^{B} \equiv E_K^{(2)} - E_{K,\pi}^{(2)} = \sqrt{U^2 + 4 J_K^2} - 2 J_K > 0.
\]

Clearly, for a given value of momentum $K$ of the dimer, 
the stronger the on-site interaction $|U|$, the smaller
the extent of the dimer wavefunction. In the case of repulsive
interaction $U > 0$ ($\alpha_K < 0$), the sign of wavefunction 
(\ref{psirdim}) alternates between the neighboring sites $i$. 
Remarkably, when $|K| = \pi/d$, and thereby $J_K = 0$, the dimer 
wavefunction is completely localized at $r_i=0$ for any $U \neq 0$. 
To illustrate the foregoing discussion, in Fig.~\ref{fig:drpwf} 
we show the dispersion relations (\ref{Ekadim}) and (\ref{Ekrdim}) 
and the wavefunctions (\ref{psiadim}) and (\ref{psirdim}) at 
$|K| = 0,\pi/2d,\pi/d$ for attractively- and repulsively-bound
dimers. 

The dimer effective mass $M^*$ for small $K$ is given by
\be
M^* = \hbar^2 \left[\frac{\partial^2 E_K^{(2)}}{\partial K^2 } \right]^{-1}_{K=0} 
= \pm \frac{\hbar^2 \sqrt{U^2 + (4 J)^2}}{4 d^2 J^2},
\ee
where the upper sign corresponds to attractive interaction $U < 0$, 
leading to positive $M^*$; while the lower sign stands for repulsive 
interaction $U > 0$, for which the effective mass $M^*$ is negative. 
For weak interaction $|U| \ll J$, we have $M^* \simeq \pm \hbar^2 /(d^2 J)
= \pm 2 m^*$, i.e., twice the single particle effective mass $m^*$ of
Eq. (\ref{effmass}). On the other hand, for strong interaction $|U| \gg J$, 
we obtain $M^* \simeq \hbar^2 /(2 d^2 J^{(2)})$, where 
$J^{(2)} \equiv - 2 J^2 /U$ is the effective tunnelling rate 
of the dimer between the neighboring lattice sites 
\cite{KWEtALPZ,molmer,PSAF,MVDP}. Now the dimer effective 
mass is large due to its slow tunneling $|J^{(2)}| \ll J$,
with the sign of $J^{(2)}$ determining the sign of  $M^*$.
In this limit the energies (\ref{Ekadim}) and (\ref{Ekrdim}) 
can be approximated as
\be
E_K^{(2)} \simeq (U - 2 J^{(2)}) - 2 J^{(2)}  \cos(K d) , \label{Ekdim}
\ee
where the first term on the right-hand side represents the dimer 
``internal energy'', while the second term is the kinetic energy 
of a dimer with quasimomentum $K$. 

%Summary
To summarize, we have considered a pair of bosonic particles in a 1D 
tight-binding periodic potential described by the Bose-Hubbard model
and derived exact analytic solutions for the two-particle problem.
These solutions describe the scattering states as well as 
interaction-bound dimer states of the two particles. We have obtained 
simple explicit expressions for the energies and wavefunctions of 
the two-particle states and outlined their properties. Our results 
are relevant to the recent experimental \cite{KWEtALPZ} and 
theoretical \cite{molmer,PSAF,MVDP} studies of strongly interacting
pairs of atoms in an optical lattice.

\end{document}